\documentclass[]{article}

\usepackage{amsmath}
\usepackage{graphicx}

\title{Equal Marginal Power for Co-Primary Endpoints}
\author{Simon Bond, Cambridge Clinical Trials Unit, \\
Cambridge University Hospitals NHS Foundation Trust, Cambridge, UK}
	
\begin{document}

\maketitle
\begin{abstract}
	The choice of sample size in the context of co-primary endpoints for a randomised trial is discussed. Current guidance can leave endpoints with unequal marginal power. A method is provided to achieve equal marginal power by using the flexibility provided in multiple testing procedures. A comparison is made to several choices of rule to determine the sample size, in terms of the study design and its operating characteristics.
\end{abstract}

\section{Introduction}

The literature on the \emph{analysis} of multiple endpoints is rich with numerous techniques to control the type 1 error rate \cite{fda_22, vickerstaff_19, hamasaki_18}. However the corresponding steps for study design, specifically choosing the sample size is less abundant.  The designation of a singular endpoint as \emph{primary} implies that the choice of sample size is justified by the properties of the future primary analysis of the same endpoint; the vast majority of clinical trials use a sample size that provides a power of 80-90\% assuming a minimally clinically important  difference, and any other nuisance parameter values.   The use of multiple endpoints as \emph{co-primary}  implies the desire to link the choice of sample size and analysis in the same way.   However rather than there just being 2 possible outcomes from one formal hypothesis test,  for $k$ co-primary endpoints there are $2^k$  combinations, and indeed  $2^k$ sets of hypotheses rather than just the pair Null and Alternative,  thus giving  $2^{2k}$  sets of considerations. 

Guidance does suggest that there may be two specific scenarios that simplify our considerations:  where \emph{all} the co-primary endpoints must achieve significance to change clinical practice;  where it would be of interest should \emph{any} of the endpoints gain significance.   These can be termed the \emph{conjunctive} and \emph{disjunctive} alternative hypotheses.  Both focus on a single specific event, and so the chances of it occurring, as an unambiguous definition of the power, can be calculated and used to inform the choice of sample size.   

Other recommendations focus on a simpler consideration of the marginal chance of each endpoint achieving significance, and thus there can be $k$ different powers or $k$ different choices of sample size unless all the input parameters for each endpoint are identical, which is highly unlikely. In which case the general guidance is to choose the largest sample size over the different endpoints. This over-powers all the other co-primary endpoints and could be considered to impose an ordering, rather than equality, on the importance of each co-primary, which generally would not be desirable and potentially unethical. 

This paper aims to improve this last case, focusing on the marginal power,  but using the extra flexibility in the analysis steps allowed by \cite{bretz_09} as a way to provide equal marginal power and sample size across a set of multiple endpoints.  We present the mathematical aspects and algorithm to perform the calculations with an example, along with some tables to potentially use to look up, rather than calculate, a suitable design for two co-primary endpoints as an important special case.  We consider the operating characteristics, across a range of specifications for two co-primary endpoints.

\section{Definitions and notation}

We assume that a randomised clinical trial wants to have a set of $k$ co-primary endpoints, each given the index $i$. Each endpoint has its own pair of minimally clinically important differences $\delta_i$ and standard deviation $\sigma_i$.  We focus on the canonical case of continuous endpoints with normally distributed z-statistics, as inference on the vast majority of endpoints used, including binary and time-to-event, will be well approximated by this distribution as justified by the central limit theorem. We take the \emph{standardised treatment effect} to be $|\delta_i|/\sigma_i.$

\section{Novel Method} \label{sec_novel}

A multiple testing procedure needs to be specified that alters the nominal significance level used in each endpoint's hypothesis test, along with a set of steps to determine which subset of endpoints has achieved statistical significance. The rationale is the need to preserve the overall family-wise error rate to a fixed level $\alpha$, typically 5\%,  under the global null hypothesis. Such a constraint can be criticized but crucially is a requirement set by regulatory bodies \cite{fda_22} for pivotal clinical trials. 

Following the general graphical approach \cite{bretz_09}, we form a graph with nodes for each endpoint and form pairs of arrows in both directions between every pairwise combination,  with equal sharing of transference of $\alpha.$  But we only constrain the initial choice of nominal significance levels $\alpha_i$  to sum to $\alpha$ the overall family-wise error rate.  At each iteration of the procedure we see if any nominal p-values are below their nominal significance level. If there are none the process stops. Otherwise we reject one of these hypotheses and transfer their nominal significance to the other hypotheses' nodes using the proportion stated on any arrows leaving the rejected node, which is deleted from the diagram. The order in  which rejected hypotheses are deleted does not matter, as proven in the original paper.

\begin{figure*}[t]
	\centering
		\includegraphics[width=\textwidth]{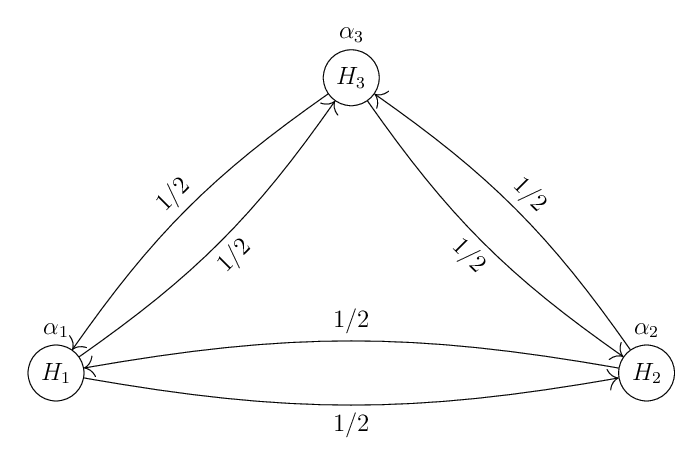}
\caption{Graphical Representation of multiple testing across three endpoints}
\label{fig_graph}
\end{figure*}

Our approach is a slight generalisation of the Bonferroni-Holm method \cite{holm_79}  as we start with unequal initial values of $\alpha_i$ rather than all equal to $\alpha/k$. Figure \ref{fig_graph} shows an example for three endpoints represented graphically.

The basic sample size formula, to achieve a marginal power  for an endpoint to be significant at the first iteration of a multiple testing procedure is 
\begin{equation}
 n= d (z_\alpha + z_\beta)^2 \sigma^2/\delta^2  \label{eqn_n} 
\end{equation}
where $d$ is a factor that solely depends on the design of the trial (parallel or crossover,  randomisation ratio), but not parameters specific to the endpoint. The $\beta$ gives the desired type II error rate,  and $z_x$ is the standard normal inverse cumulative distribution function for quantile $x$.  We conservatively approximate the marginal chances of significance by the chance that an endpoint reaches significance on the first iteration of the algorithm, and ignores subsequent iterations; under the alternative hypothesis, the probability of achieving significance at any iteration is dominated by the first iteration. 

So if we denote  $r_i =  |\delta_i / \delta_1| \sigma_1/ \sigma_i,  2 \geq i $  then we can constrain the $n$ to be equal across each end point with 

$$ z_{\alpha_i}+ z_\beta = r_i ( z_{\alpha_1}+ z_\beta ), \forall i > 1 .$$

Working on the scale of the $\{z_{\alpha_1},\ldots, z_{\alpha_k}\}' = \mathbf{z_\alpha}$ , this is simply a line in $k-$dimensional space, which can be represented in vector format
$\mathbf{z_\alpha} = \lambda \mathbf{v}+ \mathbf{v_0},$ where  $ \mathbf{v}'=\{1, r_2,\ldots, r_k\}, \mathbf{v_0}'=\{0, (r_2-1)z_\beta,\ldots, (r_k-1)z_\beta\}, $  

The other constraint is  $\sum_i \Phi(z_{\alpha_i}) = \alpha,$ so the problem reduces to a one-dimensional equation in terms of $\lambda.$ The root can be found using the Newton-Raphson method, iteratively mapping until convergence

$$\lambda \mapsto \lambda -  \left( \sum_i \Phi(z_{\alpha_i}(\lambda))- \alpha  \right) / \pmb{\phi}(\lambda)' \mathbf{v}$$

where $\pmb{\phi}(\lambda)$ has $i$th component $\phi(z_{\alpha_i}(\lambda))$,  and $\Phi$ and $\phi$ are the CDF and density functions of a standard normal. Simple application of the intermediate value theorem, proves a root exists as all the elements of $\mathbf{v}$ are positive so we can take large negative and positive values for $\lambda$ to give opposite signs to the constraint equation. 

This method provides an approximately equal marginal power for each endpoint under their individual alternative hypotheses,  and works by using the flexibility in the initial choice of nominal significance levels for each endpoint, whilst satisfying  two constraints of meeting a fixed family-wise error rate, and having a single value for the sample size as required practically when running a randomisation trial.

\section{Example}

We present an example with 4 co-primary endpoints, and use a set of $r$ values $\{1.2, 1.3, 1.5\}$ with  a power of 90\%, 1-sided FWER of 5\%, and solve for $\lambda.$ R code is supplied as supplementary material.

The resulting $\mathbf{z_\alpha}=\{ -1.78, -2.39, -2.70, -3.31\},$ which equates to $\pmb{\alpha}= \{0.0376, 0.0084, 0.0035, 0.00046\}$.   This follows the intuition that the endpoints with the bigger standardised effect size have smaller adjusted $\alpha_i$, as they would be over-powered with equal $\alpha_i$,  and inversely the under-power endpoints with smaller standardised effect sizes are boosted by having a more generous $\alpha_i$ but a relatively modest increase in sample size. 

Table \ref{tab_lookup} considers the specific case of two co-primary endpoints, which are the majority of real-life examples. It takes the cases of 1-sided $\alpha=0.05,0.025$, and power 80\% and 90\%,   across a grid of values for $r_2,$ which is now simplified in notation to $r$. The output is 
\begin{itemize}
	\item the initial $\alpha_1$ value, from which $\alpha_2$ can be obtained by subtraction
	\item the equal sample size needed, scaled by $d \sigma^2/\delta^2$ as per equation (\ref{eqn_n}).
\end{itemize}
The actual equal sample size needed will  depend on the absolute, rather than relative, values of the standardised effect size, and the nature of the trial design, as per equation (\ref{eqn_n}).

\begin{table*}[!t]
\centering

\begin{tabular*}{\textwidth}{@{\extracolsep\fill}lllll@{\extracolsep\fill}}

	$r$ &  \multicolumn{2}{c}{$\alpha=2.5\%$ }  & \multicolumn{2}{c} {$\alpha=5\%$ }\\ 
	&  Power=80\% & Power=90\% &  Power=80\% & Power=90\% \\ 
\hline
	1.1 & 1.71, 8.76&1.77, 11.46&3.26, 7.21&3.38, 9.67 \\  
	1.2 & 2.06, 8.31&2.14, 10.94&3.89, 6.79&4.06, 9.15\\
	1.3 & 2.28, 8.07&2.35, 10.68&4.34, 6.52&4.52, 8.85\\ 
	1.4 & 2.4, 7.94&2.45, 10.57&4.64, 6.36&4.78, 8.69\\ 
	1.5 & 2.46, 7.89&2.48, 10.53&4.82, 6.27&4.91, 8.62\\ 

\end{tabular*}
\caption{Splitting $\alpha$ as a function of relative standardised effect sizes for two co-primary endpoints. Number on the left is $\alpha_1$, and on the right is $n\delta^2/(d \sigma^2)$}
\label{tab_lookup}

\end{table*}

\section{ Conjunctive and Disjunctive power}

The same concept of optimally splitting up the $\alpha$ unequally to minimise the sample size needed to reach a given power for conjunction or disjunctive definitions of power can be applied. The R package mvtnorm provides a quantile function qmvnorm, that calculates equi-coordinate quantiles of a multivariate normal with arbitrary mean and covariance matrix, where the event can either be the upper tail or lower tail. 

The disjunctive power can be calculated exactly by only considering the first iteration of the multiple testing process,  as if all endpoints are non-significant initially then we have type II error under the alternative hypothesis, and so the power is the probability of the converse. So assuming, without loss of generality, the direction of the alternative hypothesis is positive,  we find the lower tail equi-coordinate quantile for $\beta$,  with mean $\mathbf{z_\alpha} \mathbf{\sigma}/\mathbf{\delta}$ and covariance of the endpoints' correlation matrix with rows and columns scaled by $\mathbf{\sigma}/\mathbf{\delta}$. It can easily be shown that  this quantile equates to $-\sqrt{n}.$

For the conjunctive power, the region of all endpoints being significant is the upper corner of the sample space, which is the region where all are significant on the first iteration,  plus small regions with both upper and lower bounds on all but one of the endpoints.  As an approximation, and for feasibility of computing, we just look at the first iteration.  Here a very similar quantile calculation is needed,  but with the upper tail used, and the quantile being for $1-\beta.$

Minimising $n$ over the choice of the $z_{\alpha_i}$  is approached using numerical optimisation routines that include the constraint  $\sum_i \Phi(z_{\alpha_i}) = \alpha.$  The R package nloptr  provides tools to do this based on \cite{NLopt} and the NLOPT\_LN\_COBYLA method  \cite{COBYLA} converged without the need for functions to calculate derivatives. 

The conjunctive and disjunctive power calculations do require the correlation matrix to be specified as an assumption, which is a weakness compared to the marginal power approach, as there may be little to no evidence to inform such values at the time of designing a study. 

\section{Operating Characteristics}
\subsection{Sample Size}

\begin{figure*}[t]
	\centering
\includegraphics[width=\textwidth]{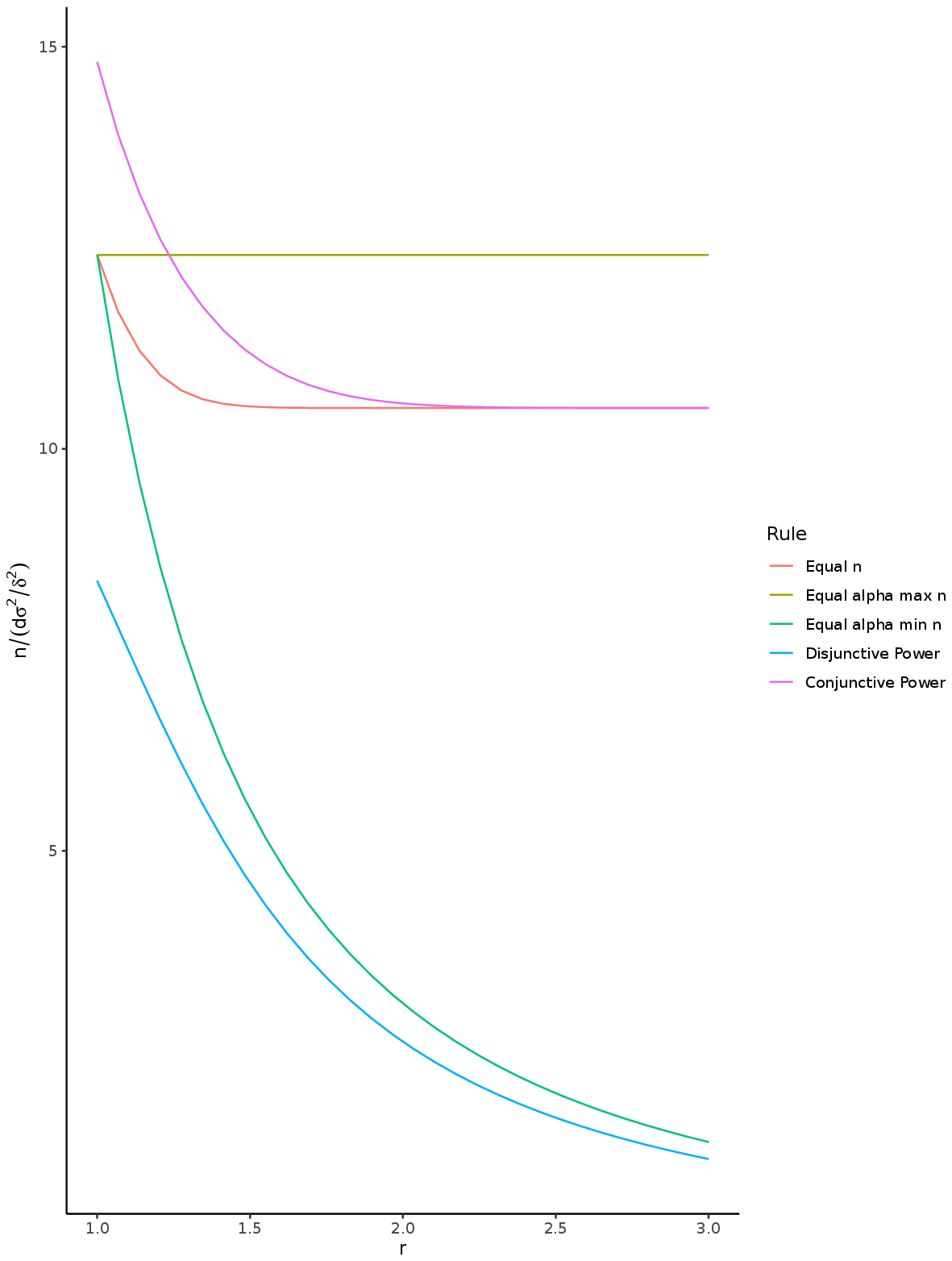}
	\caption{Comparing the sample sizes}
	\label{n_comparison}
\end{figure*}

Figure \ref{n_comparison} shows the samples size, scaled by $d \sigma^2/\delta^2,$ from different rules for $k=2$ as the ratio of effect sizes varies: using an equal marginal power and n across co-primaries; showing the maximum and minimum n from each endpoint for an equal $\alpha.$; using conjunctive and disjunctive power assuming a correlation of $0.3$.  

It shows that the equal power rule reaches an asymptote for effect size ratios larger than around 1.5. The gap between the smaller and larger n from equal alpha however keeps on growing with the endpoint with the larger effect size being over-powered. The disjunctive power requires the smallest sample size, as it is dominated by the endpoint with the largest effect size, and also adds in the chance of other endpoints reaching significance as well. As the ratio increase this dominance grows further, and is paralleled in terms of the alpha being split mostly to the endpoint with the larger standardised effect size. The conjunctive power initially starts with the highest power, as it is the most stringent combination of events, but decreases and then converges with the equal n rule based on the marginal powers as per section \ref{sec_novel}.  Both the conjunctive and marginal power is splitting the alpha mostly to the endpoint with the smaller standardised effect size	.  Overall the equal marginal power approach reaches a good compromise between the rules considered.

\subsection{Power under alternative hypotheses}

\begin{figure*}[t]
	\centering
	\includegraphics[width=\textwidth]{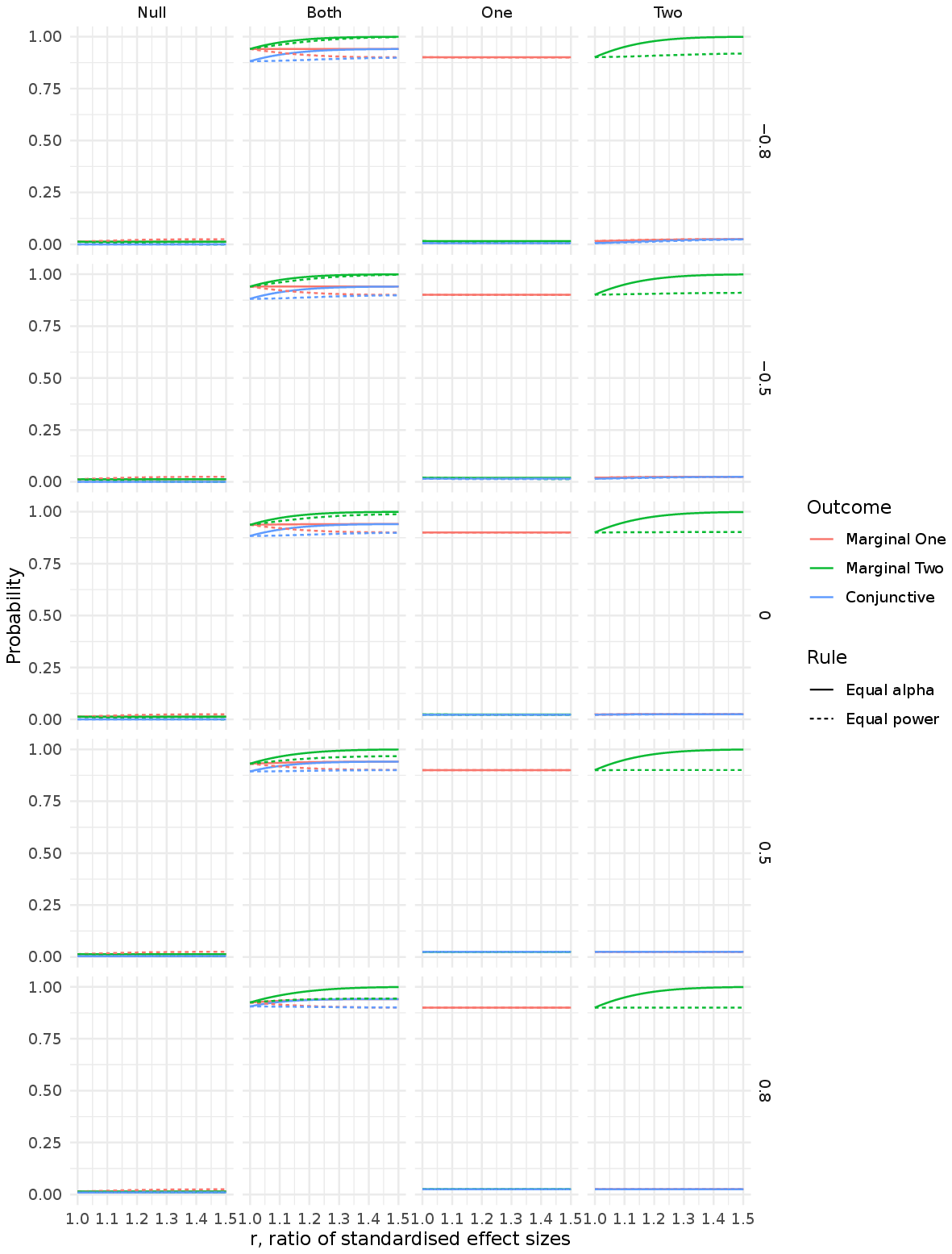}
	\caption{Comparing the probability of each possible outcome from testing}
	\label{operating_char}
\end{figure*}

The objective of this section is to convey the chances of the different possible outcomes from hypothesis testing, using exact calculations, comparing the equal alpha and equal power rules.  The parameters outside of the control of the study design are the relative effect sizes, and the correlation.  We are considering the specific case of two co-primary endpoints as this is the most common case,  and also the number of possible outcome combinations grows too large to provide any general insight for three or more endpoints. On this last point, even with  a sample space of  four mutually exclusive outcomes , in terms of which of the two endpoints achieve statistical significance (neither, both, exclusively one endpoint), the associated sigma-algebra of events is size 16. So we have simplified to only present the marginal event for each endpoint (the union of both achieving significance and just that endpoint), and both together labelled as "conjunctive". Most other events of interest could be quantified using basic probability calculus.

Considering figure \ref{n_comparison} we can see that the rule based on the conjunctive power starts as the largest n, falls below the equal alpha rule, and eventually converges to the equal power rule. Hence there is little insight to be gained from including it in this exercise.  The disjunctive power and using the smallest sample size across endpoints using equal alpha, all focus the power on the endpoint with the largest effect size, and would under-power all other endpoints, so these are also not considered, as they would be of little practical value.

Figure \ref{operating_char} is showing a large amount of information
\begin{itemize}
	\item The columns of the grid are showing four combinations of pairs of treatment effect values:
	\begin{itemize}
		\item \emph{Null} where both are zero
		\item \emph{Both} where the alternative hypothesis holds for both endpoints
		\item \emph{One} and \emph{Two} where the alternative hypotheses holds for one endpoint and the null for the other.  \emph{One} is the endpoint with the smaller effect size, and \emph{Two} is the endpoint with the larger effect determined by the ratio parameter $r$.
	\end{itemize}
	
	\item the rows of the grid show different values of the correlation between the two endpoints, going from -0.8 to +0.8.
	\item The colours of the lines represent an outcome from the multiple testing: marginally gaining significance for each endpoint,  and both endpoints simultaneously reaching significance. 
	\item the type of line, dashed or solid  tells us how the sample size and splitting of the total $\alpha$ was achieved, using a target power of 90\% and 1-sided 2.5\% family-wise significance.
	\item the horizontal axis gives $r$ the ratio of the effect size between the two endpoints
	\item the vertical axis gives a probability of the outcome. There is a black reference line at 0.9, to compare to the target power. 
\end{itemize}

The probabilities are calculated using the bivariate normal, evaluating numerically the probabilities of the bivariate z-statistic falling in the rectangular regions corresponding to the multiple testing outcomes and choice of $\mathbf{\alpha}.$  R code is supplied in supplementary material. 

Examining figure \ref{operating_char} we conclude various key points. 
\begin{itemize}
\item The probabilities reach an asymptote at around a ratio of 1.5, or are flat. 
\item The correlation has only a minor effect with small differences only visible in the \emph{Both} column where the alternative hypothesis holds for both endpoints. 
\item The type 1 error probability is well controlled whenever the strong or weak nulls hold, thus confirming the  properties of the multiple testing rules, which are not depending on sample size.  
\item When only the endpoint with the smaller effect size, \emph{One} is non-zero,  the power properties are similar for the two sets of rules for sample size and $\alpha$. However for the equivalent for the larger \emph{Two},  the equal $\alpha$ is over-powered, with an unnecessary larger sample size but the equal power is only slightly in excess of the target power.  
\item When both endpoints are non-zero, the chance of both achieving significance is close to the target power, and thus the marginal powers are in excess of the target power. 
\end{itemize}

\section{Conclusions}

The method provides a practical and readily applicable tool for the choice of sample size when using co-primary endpoints and desire is to achieve equal marginal power across the endpoints when the standardised effect size differ between endpoints. The resulting choice of sample, in the case of two co-primary endpoints only varies between a ratio 1 to 1.5 between the effect sizes, and is near constant for higher ratios,  as seen in figure \ref{n_comparison}.  The operating characteristics are good and successfully avoid the endpoints with the larger effect sizes being over-powered as would be the case when using a equal-alpha Bonferroni-Holm method for multiple adjustment and taking the maximum sample size.

\bibliography{ss_coprimary_bib}
\bibliographystyle{plain}

\end{document}